\documentclass[a4paper]{amsart}

\usepackage{amssymb,amsmath,graphicx}
\newcommand{\ii}{\mathrm{i}}

\newcommand{\dd}{\mathrm{d}}
\newcommand{\pd}{\partial}
\newcommand{\hh}{\mathcal{H}}

\newcommand{\e}{\mathrm{e}}
\newcommand{\ket}[1]{\left|#1\right\rangle}
\newcommand{\bra}[1]{\left\langle #1\right|}

\newcommand{\const}{\mathrm{const}}

\newcommand{\tr}{\mathop{\mathrm{tr}}}

\newcommand{\I}{\mathbb{I}}
\begin{document}

\title[Dynamics of Vacua\dots]{Some Notes Concerning the Dynamics
of Noncommutative Lumps Corresponding to Nontrivial Vacua in the
Noncommutative Yang--Mills Models which are perturbative branches
of M(atrix) Theory}%
\author{Corneliu Sochichiu}%
\address{Physics Dept\\ P.O. Box 2208\\ University of Crete\\ GR-71003,
Heraklion\\ GREECE}
\address{Institutul de Fizic\u a Aplicat\u a
A\c S, str. Academiei, nr. 5, Chi\c sin\u au, MD2028
MOLDOVA}%
\address{Bogoliubov Laboratory of Theoretical Physics\\
Joint Institute for Nuclear Research\\ 141980 Dubna, Moscow Reg.\\
RUSSIA}
\email{sochichi@physics.uoc.gr, sochichi@thsun1.jinr.ru}%

\thanks{Work supported by RFBR grant \#99-01-00190, INTAS grants
\#1A-262 and \#99 0590, Scientific School support grant \# 00-15-96046
and Nato fellowship program}%
%\subjclass{}%
%\keywords{}%

%\date{}%
%\dedicatory{}%
%\commby{}%
% ----------------------------------------------------------------
% ----------------------------------------------------------------
\begin{abstract}
We consider a pair of noncommutative lumps in the noncommutative
Yang--Mills/M(atrix) model. In the case when the lumps are
separated by a finite distance their ``polarisations'' do not
belong to orthogonal subspaces of the Hilbert space. In this case
the interaction between lumps is nontrivial. We analyse the
dynamics arisen due to this interaction in both naive approach of
rigid lumps and exactly as described by the underlying gauge
model. It appears that the exact description is given in terms of
finite matrix models/multidimensional mechanics whose
dimensionality depends on the initial conditions.
\end{abstract}
\maketitle
% ----------------------------------------------------------------
\section{Introduction}

Recent progress in theories over noncommutative spaces (for a
review see e.g.
\cite{Connes:2000by}--\nocite{Konechny:2000dp,Nekrasov:2000ih}%
\cite{Harvey:2001yn} and references therein), is stimulated by
their importance for the nonperturbative dynamics of string theory
\cite{Ho:1997jr}--\nocite{Connes:1998cr,Ho:1998yk}\cite{Seiberg:1999vs}.

The noncommutative models share some common features with their
commutative counterparts, however, there is a striking
dissimilarity between them in some other aspects. One particular
feature of noncommutative field theories discovered recently and
which attracted a considerable interest is that in noncommutative
models there exists a kind of localised solutions nonexistent in
the models on commutative spaces. Although they are different from
what is a \emph{soliton\/} in usual sense, these solutions are
conventionally called ``noncommutative solitons''. In actual work
we consider a subclass of such configurations. As it appears that
the ``noncommutative solitons'' in actual work even do not carry
energy (at rest) more adequate would be the term of ``vacuum'' or
``lump'' solution, throughout this paper we will keep the last
name for them.

Noncommutative lumps, in a scalar model with a potential having
nontrivial local minima were first discovered in
\cite{Gopakumar:2000zd}, in the limit of strong noncommutativity.
These solutions were interpreted as condensed lower dimensional
branes living on a noncommutative brane
\cite{Dasgupta:2000ft,Harvey:2000jt}. They were further
generalised to the case of a mild noncommutativity by allowing the
presence of the nontrivial gauge field backgrounds
\cite{Sochichiu:2000rm}--\nocite{Gopakumar:2000rw,Aganagic:2000mh}%
\cite{Harvey:2000jb}. This solutions correspond to nontrivial
gauge field fluxes
\cite{Polychronakos:2000zm}--\nocite{Bak:2000ac}\cite{Bak:2000im}.
The particular property of lump solutions we are considering in
the actual paper is that they are ``made'' purely of the gauge
fields. However using the equivalence between different
noncommutative Yang--Mills--Higgs models
\cite{Sochichiu:2000bg,Sochichiu:2000kz}, these configurations can
be mapped into noncommutative solitons in the sense of Ref.
\cite{Aganagic:2000mh} or others.

The general multi-lump solutions look like sums of projectors to
mutually orthogonal finite-dimensional  subspaces of the Hilbert
space. If subspaces are not orthogonal  the configuration fails to
be a static solution and lumps start to interact.

An approach to describe interacting lumps were proposed in
\cite{Gopakumar:2001yw,Hadasz:2001cn} by using the substitution of
the configuration of shifted lumps by a close one but being a
static solution. In this approach the interaction of the lumps is
described by the motion in the curved moduli space of static
solutions. This approach, however, would be valid only provided
that the motion is confined to the moduli space of the static
solution which requires it to be stable. There are, however,
indications that the noncommutative lumps are not stable
dynamically \cite{Acatrinei:2001xv} which leads also to the
instability of the motion around the moduli space.

Our approach is free of these drawbacks since we do not make any
assumptions about the stability. As the analysis shows the
dynamics of the system does not look as a stable one, moreover, it
appears to be stochastic! The regular motion occurs only when the
distance between lumps is exactly $\sqrt{\theta\ln 2}$. It is
interesting to note that for some natural initial conditions the
dynamics of noncommutative lumps is described by finite
dimensional matrix model.

The plan of the actual paper is as follows. First, we introduce
the reader to the noncommutative lumps in Yang--Mills--Higgs
model. After that we analyse the lump dynamics in both naive
approach when we treat lumps as rigid particles and neglect the
dynamics of the ``shapes'' of  the lumps and in an exact approach
when all possible deformations are taken into account. The
comparison reveals unexpected features in the behaviour of the
interacting lumps. We also give directions to generalise the
description to the case of lumps with arbitrary polarisations.
% ----------------------------------------------------------------

% ----------------------------------------------------------------
\section{The Model}

In this paper we consider the noncommutative gauge model described by the
following action,
\begin{equation}\label{action}
  S=\int\dd t\tr\left(\frac{1}{2}\dot{X}^i\dot{X}^i+\frac{1}{4g^2}
  [X^i,X^j]^2\right).
\end{equation}

Here fields $X^i$, $i=1,\dots,D$ are time dependent Hermitian
operators, acting on Hilbert space $\hh$ which realises a irreducible
representation for the one-dimensional Heisenberg algebra generated
by,
\begin{equation}\label{heis}
  [x^1,x^2]=\ii\theta.
\end{equation}

Operators $x^\mu$ satisfying the algebra (\ref{heis}) are said also to
be the coordinates of a noncommutative two-dimensional plane. In this
interpretation the operators of the Heisenberg algebra $\hh$ can be
represented through ordinary functions given by their Weyl symbols. The
composition rule for the symbols is given by the Moyal or star
product,
\begin{equation}\label{star}
  f*g(x)=\left.\e^{\ii\theta\epsilon^{\mu\nu}\pd_\mu\pd'_\nu}
  f(x)g(x')\right|_{x'=x},
\end{equation}
where $f(x)$ and $g(x)$ are Weyl symbols of some operators,
$f*g(x)$ is the Weyl symbol of their product and $\pd_\mu,\pd'_\mu$
denotes the derivatives with respect to $x^\mu$ and ${x'}^\mu$.
Integration of a Weyl symbol corresponds to $2\pi\theta\times$trace of
the respective operator, while the partial derivative derivative with
respect to $x^\mu$ corresponds to the commutator,
\begin{equation}
  \pd_\mu f(x)=\ii(p_\mu *f-f*p_\mu)(x)=[p_\mu,f](x),
\end{equation}
where $p_\mu$ is given by $p_\mu=(1/\theta)\epsilon_{\mu\nu}x^\nu$.
Since there is one-to-one correspondence between operators and their
Weyl symbols we will not distinguish between them, i.e. keep the same
character for both, unless in the danger of confusion.

The model (\ref{action}) corresponds to the Hilbert space
($N\to\infty$ limit) of the BFSS Matrix Model as well as in
different perturbative limits it describes the noncommutative
Yang--Mills(--Higgs) model in the temporal gauge
$A_0=0$\footnote{One should take care that the Gauss law
constraints are satisfied as well. We postpone the discussion on
the Gauss law constraints until the subsection \ref{Gauss_l}}
\cite{Sochichiu:2000ud,Sochichiu:2000bg,Sochichiu:2000kz}.

Indeed, for equations of motion corresponding to the action \eqref{action},
\begin{equation}\label{EqM}
  \ddot{X}_i+\frac{1}{g^2}[X_i,[X_i,X_j]]=0,
\end{equation}
one may find static classical solution $X_i=p_i$,
\cite{Sochichiu:2000ud,Sochichiu:2000bg,Sochichiu:2000kz},
satisfying,
\begin{equation}\label{ppp}
  [p_i,p_j]=\ii \theta_{ij}^{-1},
\end{equation}
with constant invertible $\theta_{ij}^{-1}$. We assume about the
solution also the irreducibility condition: from $[p_i,F]=0$ with
all $p_i$, $i=1,\dots,D$ it follows that $F$ is a $c$-number,
$F\sim \I$.

Expanding fields around this solution, $X_i=p_i + A_i$, and Weyl
ordering operators $A_i$ with respect to $x^i=\theta^{ij}p_j$ one
gets precisely the $(D+1)$-dimensional noncommutative Yang--Mills
model for the field given by the Weyl symbol $A_i(x)$.

Getting another solution with a smaller number of independent
$p_i$'s, $X_\alpha=p_\alpha$, $\alpha=1,\dots,p$,
\begin{equation}\label{pp}
  [p_\alpha,p_\beta]=\ii \theta_{\alpha\beta}^{-1},
\end{equation}
and $X_i=\const$, $i=p+1,p+2,\dots,D=0$ one gets as a result the
model of $p$-dimensional Yang--Mills field interacting with
($D-p$) scalars.

Having in mind this equivalence, in what follows we will consider
two-dimensional form of this noncommutative model. If one forgets
for a while also the issues with the Gauss law the theory looks
like a ``simple'' noncommutative scalar model in (2+1) dimensions.

For our purposes it will be convenient to use two-dimensional
``complex coordinates'' given by oscillator rising and lowering
operators\footnote{For the Weyl symbols we will use later $z$ and
$\bar{z}$ instead of $a$ and $\bar{a}$ to distinguish them from
the Hilbert space operators.} $a$ and $\bar{a}$,
\begin{equation}
  a=\frac{1}{\sqrt{2\theta}}(x^1+\ii x^2),\qquad
  \bar{a}=\frac{1}{\sqrt{2\theta}}(x^1-\ii x^2),\qquad
  [a,\bar{a}]=1,
\end{equation}
and the oscillator basis,
\begin{equation}\label{osc}
  \bar{a}a\ket{n}=n\ket{n},\qquad a\ket{n}=\sqrt{n}\ket{n-1},
  \qquad \bar{a}\ket{n}=
  \sqrt{n+1}\ket{n+1}.
\end{equation}

As one can see the solution \eqref{ppp} or \eqref{pp} has
divergent traces. Another type of static solutions one can find in
the model \eqref{action} is given by a configuration with
localised i.e. lump-like Weyl symbols (in some background
$p_i$).\footnote{Fairly speaking these solutions are localised if
the fields are treated as scalar ones. Since the gauge field
definition $A_\alpha=X_\alpha-p_\alpha$ implies substraction of a
linear function $p_\alpha$ this type of solutions correspond to
functions with linear growth.}
It is given by commutative matrices
of finite ranks \cite{Aganagic:2000mh}. Although these lumps carry
no energy,
--- they are geometrically nontrivial vacua, we will call them
noncommutative lumps due to their close relation to ones discussed
in the literature \cite{Gopakumar:2000zd}--\cite{Harvey:2000jb}.

Up to a gauge transformation the $N$-lump solution is given by,
\begin{equation}\label{solutionX}
  X^i=\sum_{n=0}^{N}c^i_n\ket{n}\bra{n},
\end{equation}
where $c^i_n$ is $n$-th eigenvalue of the (finite rank) operator
$X^i$. Due to the finiteness of the rank the Weyl symbol of $X^i$
vanishes at infinity as quick as Gaussian factor times a
polynomial. The simplest one-lump solutions can be written in the
form,
\begin{equation}\label{1solit}
  X_i^{(0)}=c_i\ket{0}\bra{0},
\end{equation}
where $c_i$ give the ``height'' and the ``orientation'' of the
lump, by a proper Lorentz transformation, $X_i\to\Lambda_i^jX_j$
it can be made of the form $c_i=c\delta_{i1}$, while its
``polarisation'' corresponds to the oscillator vacuum state
$\ket{0}$.

In the star-product form operator \eqref{1solit} is represented by the Weyl
symbol
$$
  X_i(\bar{z},z)=2c_i\e^{-2|z|^2}.
$$
The lump shifted along noncommutative plane by a (c-number) vector
$u$ is given by
\begin{equation}\label{1solitu}
  X_i^{(u)}=c_i\e^{-\ii p_\mu u^\mu}\ket{0}\bra{0}
  \e^{\ii p_\mu u^\mu}=c_i\e^{- |u|^2}
  \e^{\bar{a}u}\ket{0}\bra{0}\e^{-a\bar{u}}.
\end{equation}
Its Weyl symbol, correspondingly, is given by
$X_i^{(u)}(z)=2c_i\e^{-|z-u|^2}$. The shifted lump with constant
$u$ is a solution again. When $u$ becomes time-dependent one can
perform the time-dependent gauge transformation to get
\footnote{Remember the gauge: $A_0=0$.},
\begin{equation}\label{shift}
  X_i\to\e^{\ii p_\mu
  u^\mu}X_i \e^{-\ii p_\mu u^\mu},
\end{equation}
which shifts the lump back to the centre, but produce a kinetic
term for $\sim\dot{u}^2/2$. Thus a single noncommutative lump
moves freely like a \emph{non-relativistic} particle. It is also
stable since its energy at rest is zero.

In what follows we are going to analyse the situation when there
is a couple of lumps separated by a distance $u$.
% ----------------------------------------------------------------
\section{A pair of interacting lumps}

As we have shown in the previous section, a single noncommutative
lump can be always rotated to have the polarisation $\ket{0}$ and
orientation along $X_i$. When there are just two lumps one can
choose \emph{without loss of generality\/} the configuration to
involve nontrivially only two matrices e.g. $X_1$ and $X_2$.

Consider two lumps which are obtained from $c\ket{0}\bra{0}$ by
shifts along the noncommutative plane by respectively $u_{(1)}$
and $u_{(2)}$. Since the dynamics of the centre is free and can be
decoupled by a time-dependent gauge transformation similar to
\eqref{shift}, where $u=u_{(1)}+u_{(2)}$ is the coordinate of the
centre.

Thus, the configuration we consider looks like,
\begin{subequations}\label{2solit}
\begin{align}\label{2solit1}
  &X_1=c_1VPV^{-1}\equiv c\ket{-u/2}\bra{-u/2},\\ \label{2solit2}
  &X_2=c_2V^{-1}PV \equiv c\ket{u/2}\bra{u/2}\\
  &X_i=\const,\qquad i=3,\dots,D,
\end{align}
\end{subequations}
where we introduced the shorthand notations,
\begin{align}\label{V}
  & V=\e^{(\ii/2) p_\mu u^\mu}=
  \e^{\frac{1}{2}(a\bar{u}-\bar{a}u)},\\ \label{P}
  & P=\ket{0}\bra{0}.
\end{align}
The quotient $c$ can be absorbed by the rescaling of the coupling
and the time, therefore we can put it generically to unity $c=1$.

In what follows also $X_i$ $i=3,\dots,D$ will enter trivially in
the equations, so in the remaining part of the paper for the
simplicity of notations the index $i$ will run the range $i=1,2$.
If we were considering more than two noncommutative lumps we would
have to keep more matrices.

% -----------------------------------------------------------------
\subsection{Naive picture: rigid lumps}
Consider first a naive approach where we are dealing with rigid
interacting lumps which means that we are neglecting the
deformations in their shapes. In this case the only parameter
which is dynamical is the separation distance $u$. Although, this
approximation sounds reasonable, later we will consider the exact
description which shows that this approach is not justified.
However, we decided to keep this naive analysis for illustrative
purposes.

To obtain the action describing the dynamics let us insert the
ansatz \eqref{2solit} into the classical action (\ref{action}).
The computation of derivatives and traces gives for the following
$u$-dependence of the action,
\begin{equation}\label{u_action}
  S[u]=\int \dd t\left(\frac{1}{2\theta}|\dot{u}|^2-2\e^{-\frac{|u|^2}{2\theta}}
  (1-\e^{-\frac{|u|^2}{2\theta}})\right),
\end{equation}
where we restored the explicit $\theta$ dependence.
% ----------------------------------------------------------------
\begin{figure}
   \includegraphics[width=50mm]{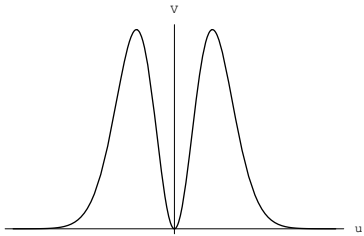}
   \caption{The profile of the lump-lump interaction potential
   in the naive approach.}
   \label{Pot}
\end{figure}
% ----------------------------------------------------------------

The potential is depicted on the fig.\ref{Pot}. According to it
sufficiently close lumps attract while distant ones repel. At the
critical distance $u_c=\sqrt{2 \theta\ln 2 }$ they will stay in
unstable equilibrium.

The above conclusions concerning the lump dynamics would be valid,
however, only in the case when one can neglect the involvement of
the lump shape in the dynamics. To evaluate the importance of the
shape dynamics one should consider arbitrary deformations of the
shape of lumps and separate them from the motion of the lump as a
whole.

In the next subsection we analyse the dynamics from the point of
view of exact field equations of motion. The lump configuration is
taken to be the initial condition for the field equations. The
result we obtain in the next section will invalidate the results
of the actual naive approach, however, the critical distance $u_c$
will correspond to a special case.
% -----------------------------------------------------------------
\subsection{Exact description: Lumps at rest.}
The exact description of the lump dynamics is given by the field
equations of motion for $X_i$,
\begin{equation}\label{em}
  \ddot{X}_i+\frac{1}{g^2}[X_j,[X_j,X_i]]=0
\end{equation}
corresponding to the action (\ref{action}), supplied with initial
conditions given by the lump background \eqref{2solit}. Since the
equations are second order, in addition to this one has to
consider the initial values for the time derivatives of $X_i$. The
simplest choice is when the configuration at $t=0$ is static.
Thus, the initial conditions we impose are as follows,
\begin{subequations}\label{ic_true_no_speed}
\begin{align}
  X_1|_{t=0}&=\ket{-u/2}\bra{-u/2},\qquad &\dot{X}_1|_{t=0}=0,\\
  X_2|_{t=0}&=\ket{u/2}\bra{u/2},\qquad &\dot{X}_2|_{t=0}=0  .
\end{align}
\end{subequations}

Considering the lumps in the initial moment as being at rest,
produces a considerable simplification to the equations of motion.
Indeed, the initial data \eqref{2solit} imply that the operators
$X_i$ are nonzero only on the two-dimensional subspace $\hh_u$ of
the Hilbert space which is the linear span of vectors $\ket{u/2}$
and $\ket{-u/2}$. Since, in virtue of equations of motion
\eqref{em}, the second time derivative is proportional to
commutators of $X_i$ then it also vanishes outside the
two-dimensional subspace $\hh_u$. Due to zero initial conditions
for the first derivatives operators $X_i$ will remain all the time
in the same two-dimensional subspace of the Hilbert space.

Let us consider only those components of $X_i$ which are nonzero. This
reduces the Hilbert space operators to ones acting on the two-dimensional
subspace $\hh_u$ of the Hilbert space spanned by $\ket{\pm u/2}$. Let us
introduce an orthonormal basis on $\hh_u$.

The natural orthonormal basis one can build up out of $\ket{\pm
u/2}$, is given by vectors $\ket{\pm}$, defined as follows (see
the Appendix),
\begin{subequations}\label{v+-}
\begin{align}\label{v+}
  &\ket{+}
  \equiv
   \begin{pmatrix}
   1\\
   0
   \end{pmatrix}=
  \frac{1}{\sqrt{2(1+\e^{-\frac12|u|^2})}}
  \left(\ket{u/2}+\ket{-u/2}\right)
   ,\\ \label{v-}
   &\ket{-}
   \equiv
   \begin{pmatrix}
   0\\
   1
   \end{pmatrix}
   =\frac{1}{\sqrt{2(1-\e^{-\frac12|u|^2})}}
   \left(\ket{u/2}-\ket{-u/2}\right).
\end{align}
\end{subequations}
The singularity in the $\ket{-}$ in the limit $u\to 0$ appears since
in this limit $\ket{u/2}$ and $\ket{-u/2}$ tends to be parallel and
the subspace become one-dimensional.

In this basis our problem is reformulated in terms of the $2\times
2$ matrix model with equations of motion superficially looking the
same as (\ref{em}),
\begin{equation}\label{em2x2}
  \ddot{X}_i^{(2)}+\frac{1}{g^2}[X_k^{(2)},[X_k^{(2)},X_i^{(2)}]]=0,
\end{equation}
but now $X_i^{(2)}$ are finite dimensional $2\times 2$ matrices.
The initial conditions in the basis \eqref{v+-} are rewritten as
follows,
\begin{subequations}\label{ic2x2}
\begin{align}\label{ic1}
  &X_1^{(2)}|_{t=0}=\frac{1}{2}
       \begin{pmatrix}
       1+\e^{-\frac12|u|^2}& -\sqrt{1-\e^{-|u|^2}} \\
       -\sqrt{1-\e^{-|u|^2}}&1-\e^{-\frac12|u|^2}
       \end{pmatrix},
       \qquad \dot{X}_1|_{t=0}=0;\\
       \label{ic2}
  &X_2^{(2)}|_{t=0}=\frac{1}{2}
       \begin{pmatrix}
       1+\e^{-\frac12|u|^2}& \sqrt{1-\e^{-|u|^2}} \\
       \sqrt{1-\e^{-|u|^2}}&1-\e^{-\frac12|u|^2}
       \end{pmatrix},\qquad \dot{X}_2|_{t=0}=0.
\end{align}
\end{subequations}

(It is worthwhile to note that the description in terms of
2$\times$2 matrices is valid  only for the situation where the
lumps were initially at rest. Beyond this the conditions
$\dot{X}_i|_{t=0}=0$ is an additional specification and it says
that also the shapes of the lumps are not changing at the initial
moment. One may try, however relax the last condition and consider
more general initial configurations including lumps with
rising/decreasing height for $\dot{X}_i\neq 0$. One can consider
more general initial condition $\dot{X}_i|_{t=0}\propto X_i$ for
which the same description in terms of 2$\times$2 matrices remains
valid. Solutions of this type for $u=0$ were considered in
\cite{Hashimoto:2000ys}.)

Similar equations as ones given by (\ref{em2x2}) although in a
different context of the one-dimensional ordinary Yang--Mills
model were under study for a long time and were initiated by
\cite{Baseian:1979zx}--\nocite{Matinian:1981ji}\cite{Matinian:1981dj}.
In the modern context of the application to the finite $N$ matrix
model they appear in
\cite{Banks:1997vh}--\nocite{Douglas:1997yp,Aref'eva:1997es}%
\cite{Aref'eva:1998mk}. The system described by such equations was
shown to exhibit a stochastic behaviour. Let us describe it in
more details to the application to the present case.

In order to rewrite the equations \eqref{ic2x2} in the scalar form
let us expand the matrices $X_i$ in terms of the two-dimensional
Pauli matrices $\sigma_\alpha$, $\alpha=1,2,3$, and the
two-dimensional unity matrix $\I_2$ (which in fact is the
projector to $\hh_u$) satisfying,
\begin{equation}\label{sigma}
  [\sigma_\alpha,\sigma_\beta]=
  \ii\epsilon_{\alpha\beta\gamma}\sigma_\gamma,\qquad [\sigma_\alpha,\I_2]=0,
\end{equation}
The expansion is as follows,
\begin{equation}\label{expansion}
  X_{1,2}=X_{1,2}^{0}\I_2+X_{1,2}^\alpha\sigma_\alpha.
\end{equation}

In terms of this expansion the equations of motion look as
follows,
\begin{subequations}\label{pa_matr}
\begin{align}\label{x0}
  &\ddot{X}_{1,2}^{0}=0 \\ \label{x1a}
  &\ddot{X}_1^\alpha+\frac{1}{g^2}(X_2^2\delta_\beta^\alpha-
  X_2^\alpha X_{2\beta})X_1^\beta=0 \\ \label{x2a}
  &\ddot{X}_2^\alpha+\frac{1}{g^2}(X_1^2\delta_\beta^\alpha-
  X_1^\alpha X_{1\beta})X_2^\beta=0,
\end{align}
where $X_{1,2}^2=X_{1,2}^{\alpha}X_{1,2}^{\alpha}$. For the
initial conditions one also has,
\begin{align}\label{ic_x0}
  & \dot{X}_{1,2}^\alpha=0,\\ \label{ic_x^0}
  & X_1^0|_{t=0}=X_2^0|_{t=0}=\frac12 \\ \label{ic_x^1}
  & X_1^1|_{t=0}=-\frac12\sqrt{1-\e^{-|u|^2}},\qquad
  X_2^1|_{t=0}=\frac12\sqrt{1-\e^{-|u|^2}},\\ \label{ic_x^2}
  & X_1^2|_{t=0}=X_2^2|_{t=0}=0,\\ \label{ic_x^3}
  & X_1^3|_{t=0}=X_2^3|_{t=0}=\e^{-\frac12|u|^2}.
\end{align}
\end{subequations}

In particular, the equation (\ref{x0}) says that the scalar parts of the
matrices $X_{1,2}$ remains constant during the motion ($X_{1,2}^0(t)=1/2$)
provided the zero initial conditions for the ``velocities'' $\dot{X}_i^0=0$.
At the same time the remaining parts are subjects to more complicate
nonlinear dynamics.

Before analysing the solutions for $X_{1,2}^\alpha$,
$\alpha=1,2,3$ let us consider their interpretation in terms of
the lump dynamics over noncommutative space in the star-product
representation.

The noncommutative function which corresponds to a particular solution
$X_i^\alpha(t)$ will be given by
\begin{equation}\label{star_sol}
  X_i(t;z,\bar{z})=\frac{1}{2}\I(z,\bar{z})+X_i^\alpha(t)\sigma_\alpha(z,\bar{z}),
\end{equation}
where $X_i^\alpha(t)$ are the solutions of to \eqref{pa_matr} and
$\I(z,\bar{z})$ with $\sigma_\alpha(x)$ are the Weyl symbols corresponding to
the Pauli matrices.

The respective Weyl symbols are computed in the Appendix. They are given by,
\begin{subequations}\label{sigmas}
\begin{align}\label{sigma1}
  \sigma_1(z,\bar{z})&=\frac{2}{\sqrt{1-\e^{-|u|^2}}}
  \left(\e^{-2|z-\frac{u}{2}|^2}-\e^{-2|z+\frac{u}{2}|^2}\right),\\
\label{sigma2}
  \sigma_2(z,\bar{z})&=\frac{2\ii\e^{-2\bar{z}z}}{\sqrt{1-\e^{-|u|^2}}}
  \left(\e^{\bar{z}u-\bar{u}z}-\e^{-\bar{z}u+\bar{u}z}\right),\\
\label{sigma3}
  \sigma_3(z,\bar{z})&=-\frac{2\e^{-\frac12|u|^2}}{1-\e^{-|u|^2}}
  \left(\e^{-2|z-\frac{u}{2}|^2}+\e^{-2|z+\frac{u}{2}|^2}\right)\\
  \nonumber
  & \qquad +\frac{\e^{-2\bar{z}z}}{1-\e^{-|u|^2}}
  \left(\e^{\bar{z}u-\bar{u}z}+\e^{-\bar{z}u+\bar{u}z}\right),\\
\label{sigm0}
  \I(\bar{z},z)&=\sigma_0(z,\bar{z})=
  \frac{2}{1-\e^{-|u|^2}}
  \left(\e^{-2|z-\frac{u}{2}|^2}+\e^{-2|z+\frac{u}{2}|^2}\right)\\
  \nonumber
  & \qquad -\frac{2\e^{-2|z|^2-\frac12 |u|^2}}{1-\e^{-|u|^2}}
  \left(\e^{\bar{z}u-\bar{u}z}+\e^{-\bar{z}u+\bar{u}z}\right).
\end{align}
\end{subequations}

As it can be seen from the eqs. \eqref{star_sol}, \eqref{sigmas}
irrelevant to the particular form of the solution
$X_i^{\alpha}(t)$, at any time fields $X_i(z,\bar{z})$ are nonzero
only in the small vicinities (of the size of the order of
$\sim\sqrt{\theta}$) of of points $z=0$ and $z=\pm u/2$. This
means that irrespective to the initial distance between the lumps
once left with zero initial velocities they will not try to leave
their places, the dynamics instead will concern only their heights
and creation of a ``baby-lump'' in the middle point between them.
Let us note that this should be surprising as it is in total
disagreement with the naive approach drawn in the previous
subsection, since there is no regime when the lumps would behave
like rigid particles.

Let us consider now the the time-dependent functions $X_i^\alpha (t)$  in
more details. The equations \eqref{pa_matr} are too complicate to find the
general solution, however, for our particular initial data one can use the
rich symmetry of the model and find a simplifying ansatz.

Assuming that the magnitudes of $X_1^\alpha$ and $X_2^\alpha$ are equal
$X_1^2(t)=X_2^2(t)$ also for nonzero times, we can check this assumption
later as a consistency condition for the ansatz, but also prove it
independently of the ansatz using conservation laws, one can split
$X_1^\alpha$ and $X_2^\alpha$ into two orthogonal components $X^\alpha$ and
$Y^\alpha$ as follows,
\begin{subequations}\label{ans_1}
\begin{align}\label{x12}
  &X_1^\alpha=X^\alpha+Y^\alpha,\qquad
  X_2^\alpha=X^\alpha-Y^\alpha,\\ \label{xy}
  &X=\frac12 (X_1^\alpha+X_2^\alpha),\qquad  Y=\frac12
  (X_1^\alpha-X_2^\alpha),
\end{align}
\end{subequations}
the equality of the square modules $X_1^2(t)=X_2^2(t)$ implies that $X$ and
$Y$ remain orthogonal. The equations of motion in terms of $X$ and $Y$ read,
\begin{subequations}\label{XY}
\begin{align}\label{X}
  &\ddot{X}^\alpha=-\frac{2}{g^2}Y^2X^\alpha, \\ \label{Y}
  &\ddot{Y}^\alpha=-\frac{2}{g^2}X^2Y^\alpha,
\end{align}
where $X^2=X^\alpha X^\alpha$ and $Y^2=Y^\alpha Y^\alpha$. The initial
conditions are respectively,
\begin{align}\label{ic_XY1}
  & X^\alpha|_{t=0}=\frac12 (X_1^\alpha(0)+X_2^\alpha(0)),\qquad
  Y^\alpha|_{t=0}=\frac12 (X_1^\alpha(0)-X_2^\alpha(0)),\\
  \label{ic_XY2}
  & \qquad \dot{X}^\alpha(0)=\dot{Y}^\alpha(0)=0.
\end{align}
\end{subequations}

From eqs. \eqref{XY} one can see that the directions of $X^\alpha$
and $Y^\alpha$ do not change. The fact that $X^\alpha$ and
$Y^\alpha$ are always mutually orthogonal makes the assumption
$X_1^2(t)=X_2^2(t)$ for the ansatz \eqref{ans_1} consistent.

Splitting the vectors $X^\alpha$ and $Y^\alpha$ in the dynamical
magnitude and the static direction, given by the constant unimodular
vectors
\begin{subequations}
\begin{align}\label{eX}
  &e_X^\alpha=X^\alpha/\sqrt{X^2}|_{t=0}=(0,0,1),\\ \label{eY}
  &e_Y^\alpha=Y^\alpha/\sqrt{Y^2}|_{t=0}=(1,0,0),
\end{align}
\end{subequations}
one has the equations for the magnitudes $X$ and $Y$,
\begin{subequations}\label{2d}
\begin{align}\label{X_sc}
  &\ddot{X}=-\frac{2}{g^2}Y^2X, \\ \label{Y_sc}
  &\ddot{Y}=-\frac{2}{g^2}X^2Y,
\end{align}
which are supplied by the initial data,
\begin{equation}\label{ic_XY_sc}
  X|_{t=0}= \e^{-\frac12 |u|^2},\quad
  Y|_{t=0}=-\sqrt{1-\e^{-|u|^2}},\quad \dot{X}|_{t=0}=\dot{Y}_{t=0}=0.
\end{equation}
\end{subequations}

As we mentioned earlier, the system \eqref{2d} exhibits a
stochastic behaviour which has been studied both numerically and
analytically \cite{Baseian:1979zx}--\cite{Aref'eva:1998mk}. The
system is equivalent to one of a two-dimensional particle in the
potential $U(X,Y)=X^2Y^2$. The allowed by energy conservation
region of the configuration space is divided into so called
stadium $X\sim Y \lesssim 1$ where the motion is almost free and
four channels along the axes. The motion in channels can be
described by the asymptotic formula, in the limit when one
coordinate is much smaller than another, say $X\ll Y$
\cite{Medvedev:1985ja},
\begin{subequations}\label{asy}
\begin{align}\label{as_solY}
   &Y(t)=-\frac{A}{2}t^2+W_0t+Y_0 \\ \label{as_solX}
   &X(t)=\frac{1}{g^2}\sqrt{\frac{2A}{Y(t)}}
   \cos
   \left[g\left(-\frac{A}{6}t^3+\frac{W_0}{2}t^2
   Y_0t+\varphi_0\right)\right],
\end{align}
where
\begin{equation}
  A=\frac{V_0^2+g^2 X_0^2Y_0^2}{Y_0},\qquad
  \varphi_0=\arccos \sqrt{\frac{X_0^2Y_0^2}{V_0^2+g^2X_0^2Y_0^2}}
\end{equation}
and
\begin{align}
  &X_0=X(t_0),\qquad  &Y_0=Y(t_0),\\
  &V_0=\dot{X}(t_0),\qquad &W_0=\dot{Y}(t_0),
\end{align}
$t_0$ being the time of entrance into the channel.
\end{subequations}

In the channel the particle reach the maximal value of $Y\sim
W_0^2/A$ after which it is reflected back to the stadium. The
instability arises when the particle passes through the stadium
and enters a new channel. So, generally the motion of the particle
is stochastic. Also there is a discrete set of trajectories which
are closed. Thus, depending on initial conditions the system can
move in a regular periodic way, although this motion is unstable
as arbitrary small perturbation can push the system to the
stochastic regime.

The asymptotic formulae \eqref{as_solY}, \eqref{as_solX} can
provide a reliable description of the system for a certain period
of time for extremal cases when the lump centre separation
distance is either large  ($q\equiv\e^{-|u|^2}\ll 1$) or small
($\sqrt{1-q^2}\ll 1$). Thus, if $u\to\infty$ ($q\ll 1$) then for
times less than $ t_{\mathrm{stoch}}= gq^{-1}$, one has the
asymptotic solution,
\begin{subequations}
\begin{align}\label{u->infty}
  &Y(t)=\frac{q^2\sqrt{1-q^2}}{4g^2}t^2-\sqrt{1-q^2},\\
  &X(t)=\sqrt{\frac{q^2(1-q^2)^{1/2}}{(q^2(1-q^2)^{1/2}/4g^2)
  t^2-(1-q^2)^{1/2})}}\\ \nonumber
  &\qquad \times\cos\left[\frac{1}{g}\left(\frac{
  q^2\sqrt{1-q^2}}{4g^2}t^3-\sqrt{1-q^2}t\right)\right].
\end{align}
\end{subequations}
In the opposite case when the lumps are close one can again give a
reliable description for of the dynamics by the asymptotic
formula,
\begin{subequations}
\begin{align}
  & X(t)=-\frac{q(1-q^2)}{4g^2}t^2+q,\\
  & Y(t)=\sqrt{\frac{q(1-q^2)}{-(q(1-q^2)/4g^2)
  t^2+q}}\cos\left[\frac{1}{g}\left(-\frac{
  q(1-q^2)}{4g^2}t^3+qt\right)\right],
\end{align}
\end{subequations}
valid for times up to of order $t_{\mathrm{stoch}}=
g(1-q^2)^{-1/2}$ after which the system approaches the stadium
where we cannot control it.

There is also one particular separation distance which corresponds
to periodic motion. This happens for the initial conditions
$X|_{t=0}=-Y|_{t=0}=1/\sqrt{2}$ or $u=\sqrt{\theta\ln2}$. In this
case the motion is periodic and is given by $X(t)=Y(t)\equiv f(t)$
where for $f(t)$ we have the (implicit) formula,
\begin{equation}\label{implicite}
  f(t):\qquad t=\int_{1/\sqrt{2}}^f \frac{\dd u}{\sqrt{1/4-u^4}}.
\end{equation}

Now, let us recall that in terms of $X(t)$ and $Y(t)$ the
dynamical field $X_i(t,\bar{z},z)$ describing the lumps takes
according to eq. \eqref{star_sol} the following form,
\begin{subequations}\label{sol}
\begin{multline}
  X_1(t,\bar{z},z)=\frac12
  \sigma_0(\bar{z},z)+X(t)\sigma_3(\bar{z},z)+Y(t)\sigma_1(\bar{z},z)=
  \\
  \frac{1-\e^{-\frac12
  |z|^2}X(t)-\sqrt{1-\e^{-|u|^2}Y(t)}}{1-\e^{-|u|^2}}
  \e^{-2|z-\frac{u}{2}|^2}\\
  +\frac{1-\e^{-\frac12
  |z|^2}X(t)+\sqrt{1-\e^{-|u|^2}Y(t)}}{1-\e^{-|u|^2}}
  \e^{-2|z+\frac{u}{2}|^2}\\
  +\frac{X(t)-\e^{-\frac12 |u|^2}}{1-\e^{-|u|^2}}\e^{-2|z|^2}
  (\e^{\bar{z}u-z\bar{u}}+\e^{-\bar{z}u+z\bar{u}}),
\end{multline}
and,
\begin{multline}
  X_2(t,\bar{z},z)=\frac12
  \sigma_0(\bar{z},z)+X(t)\sigma_3(\bar{z},z)-Y(t)\sigma_1(\bar{z},z)=\\
  \frac{1-\e^{-\frac12
  |z|^2}X(t)+\sqrt{1-\e^{-|u|^2}Y(t)}}{1-\e^{-|u|^2}}
  \e^{-2|z-\frac{u}{2}|^2}\\
  +\frac{1-\e^{-\frac12
  |z|^2}X(t)-\sqrt{1-\e^{-|u|^2}Y(t)}}{1-\e^{-|u|^2}}
  \e^{-2|z+\frac{u}{2}|^2}\\
  +\frac{X(t)-\e^{-\frac12 |u|^2}}{1-\e^{-|u|^2}}\e^{-2|z|^2}
  (\e^{\bar{z}u-z\bar{u}}+\e^{-\bar{z}u+z\bar{u}})
\end{multline}
\end{subequations}
where the functions $\sigma_{1,3}$ are given by the eqs.
\eqref{sigmas}. Let us note that the function
$\sigma_1(\bar{z},z)$ is localised in the points where the lumps
are i.e. at $z=\pm u/2$ while the function $\sigma_3(\bar{z},z)$
is localised at both lump positions as well as at the origin where
is the middle of the lump centres connecting line.

The analysis of the solution \eqref{sol} reveals that once left at
their positions the lumps will not tend to move away from them but
engage in a stochastic change of their heights as well as creation
of a ``baby'' lump in the middle point between them which is the
origin of the noncommutative plane. This process can be reliably
described for a while of time in the limits when the lumps are
placed very close or very far, each case degenerating to
stochastic, although correlated variation in the heights of the
lumps.

% -----------------------------------------------------------------
\subsection{Exact description: Lumps in motion.}
The difference arising when considering moving lumps consists in
the initial values for the velocities. Since a generic initial
condition for the velocities can complicate the system making it
back infinite dimensional we confine ourselves to such initial
configurations which correspond to rigid motion of the lumps.

Thus, one has to replace the initial values for the velocities by
the following,
\begin{equation}\label{speed}
   \dot{X}_i|_{t=0}=\frac{\pd X_i}{\pd u}\dot{u}|_{t=0}
   +\frac{\pd X_i}{\pd\bar{u}}\dot{\bar{u}}|_{t=0} ,
\end{equation}
or, explicitly, using  \eqref{2solit},
\begin{subequations}\label{speeds}
\begin{align}
  \dot{X}_1|_{t=0}&=-\frac14({\bar{v}}u+
  \bar{u}v)X_1|_{t=0}+\frac12(v
  \bar{a}X_1+{\bar{v}}X_1a)|_{t=0},\\
  \dot{X}_2|_{t=0}&=-\frac14({\bar{v}}u+
  \bar{u}v)X_2|_{t=0}-\frac12(v
  \bar{a}X_2+{\bar{v}}X_2a)|_{t=0},
\end{align}
\end{subequations}
where $v=\dot{u}(t=0)$, and solve the infinite dimensional
operator equation \eqref{EqM}.

Applying the same strategy as in the case of lumps at rest we see
that the initial data are given by operators which are nonzero
only in a four-dimensional subspace $\hh_u^v$ of the
infinite-dimensional Hilbert space $\hh$, which is spanned by our
two old vectors $\ket{\pm u/2}$ together with other two newcomers
$\bar{a}\ket{\pm u/2}=2(\pd/\pd u\ket{\pm u/2})$. Let us note that
they are all linear independent for $u\neq 0$. Therefore, the
system is reduced to the four-dimensional matrix model.

The treatment of this model differs from what we had with lumps at
rest only in technical details, therefore we will not discuss it
more.

The qualitative picture one has in this situation does not change
much in comparison to the case of lumps a rest. Just as in the
previous case there is a stochastic dynamics of the heights of the
lumps and creation of ``baby''-lumps while the centres of the
lumps will keep moving with the constant velocities. Indeed, for
accelerating lump operators $X_i$ are nonzero out of the subspace
$\hh_u^v$, which, as we know, does not happen.

In general, the solution is given by a linear combination with
time dependent coefficients of functions \eqref{symbols}, their
first derivatives $(\pd \sigma_\alpha /\pd u)$, $(\pd\sigma_\alpha
/\pd \bar{u})$ and some of their second derivatives like
$(\pd^2\sigma_\alpha/\pd u\pd\bar{u})$.

\subsection{The Gauss Law}\label{Gauss_l}
Once we want to relate our system to the the Yang--Mills/BFSS
model we have to care about the Gauss law constraint, which is
obtained from the variation of the $A_0$ component of the original
gauge invariant noncommutative Yang--Mills or BFSS action. This
constraint looks like follows,
\begin{equation}\label{gl}
   L=[X_i,\dot{X}_i]=0.
\end{equation}

As we discussed at the beginning of this section the equations of
motion imply that the quantity \eqref{gl} is at least
conserved. Indeed, using the equations of motion one has
\begin{equation}\label{conserv}
 \dot{L}=\ii[X_i,\ddot{X}_i]=0.
\end{equation}

Therefore to get a self-consistent solution for the
Yang--Mills/M(atrix) theory one has to verify that $L|_{t=0}=0$.
For zero velocity initial conditions this is implied
automatically, while for the moving lumps one has,
\begin{equation}\label{gl_speeds}
  L=\ii[X_i,\dot{X}_i]|_{t=0}=\ii v\left[\left(\frac{\bar{u}}{2}-
  \bar{a}\right)X_1+\left(\frac{\bar{u}}{2}+
  \bar{a}\right)X_2\right]+\text{h.c.},
\end{equation}
where ``h.c.'' stands for the Hermitian conjugate.

The equation \eqref{gl_speeds} requires velocities $v$ and
$\bar{v}$ to vanish. However, the violation of the Gauss law for
nonzero velocities may be interpreted as the presence of
nontrivial electric charge. Indeed, in the presence of external
sources the Gauss law becomes,
\begin{equation}\label{gl_2}
  L'=\ii[X_i,\dot{X}_i]+\rho=0,
\end{equation}
where $\rho$ is some electric charge density which appears in
the action as a term $\Delta S_{\mathrm{charge}}=\int \dd^{p+1}x\,
\rho X_0$ and which is chosen
to cancel \eqref{gl_speeds} exactly. (Here we are not going to analyse
in which conditions such charge density can be created.)

As a result we have that the Gauss low is satisfied automatically
in the case of the lumps at rest, while moving lumps generate some
background charge distribution.

\subsection{More dimensions}
One can do analogous analysis in more than $2+1$ dimensions.

The only difference which appear in $p+1$ dimensions is that one
has to compute the Weyl symbols of sigma matrices with respect to
a different background e.g. one given by \eqref{ppp}. As a result
one has equations similar to \eqref{sigmas},
\begin{subequations}\label{sigmas-multi}
\begin{align}\label{sigma1m}
  \sigma_1(x)&=\frac{2}{\sqrt{1-\e^{-\frac{1}{2}u\cdot G\cdot u}}}
  \left(\e^{-(x-\frac{u}{2})\cdot G\cdot (x-\frac{u}{2})}-
  \e^{-(x+\frac{u}{2})\cdot G\cdot (x+\frac{u}{2})}\right),\\
\label{sigma2m}
  \sigma_2(x)&=-\frac{4\ii\e^{-x\cdot G \cdot x}}
  {\sqrt{1-\e^{-\frac{1}{2}u\cdot G\cdot u}}}
  \sin u\times x,\\
\label{sigma3m}
  \sigma_3(x)&=-\frac{2\e^{-u\cdot G\cdot u}}
  {1-\e^{-\frac{1}{2}u\cdot G\cdot u}}
  \left(\e^{-(x-\frac{u}{2})\cdot G\cdot (x-\frac{u}{2})}
  +\e^{-(x+\frac{u}{2})\cdot G\cdot (x+\frac{u}{2})}\right)\\
  \nonumber
  & \qquad +\frac{2\e^{-x\cdot G\cdot x}}{1-\e^{-\frac{1}{2}u\cdot G\cdot u}}
  \cos u\times x,\\
\label{sigm0m}
  \I(x)&=\sigma_0(x)=
  \frac{2}{1-\e^{-\frac{1}{2}u\cdot G\cdot u}}
  \left(\e^{-(x-\frac{u}{2})\cdot G\cdot (x-\frac{u}{2})}+
  \e^{-(x+\frac{u}{2})\cdot G\cdot (x+\frac{u}{2})}\right)\\
  \nonumber
  & \qquad -\frac{4\e^{-x\cdot G\cdot x-\frac14
  u\cdot G\cdot u}}{1-\e^{-\frac{1}{2}u\cdot G\cdot u}}\cos u\times x,
\end{align}
\end{subequations}
where we introduced the notation, $u\times x=\theta_{\mu\nu}u^\mu
x^\nu$, $\mu,\nu=1,\dots,p$, and squares are computed with the
metric $G=+\sqrt{-\theta^{-2}}$. In the basis for which the
non-commutativity matrix $\theta^{\mu\nu}$ takes the canonical
form,
\begin{equation}\label{canon}
  \theta^{\mu\nu}=
  \begin{pmatrix}
  \theta_{(1)}\ii \sigma_2&0&0&\hdots \\
  0&\theta_{(2)}\ii \sigma_2&0&\hdots\\
  0&0&\theta_{(3)}\ii \sigma_2&\hdots\\
  \vdots&\vdots&\vdots&\ddots
  \end{pmatrix},
\end{equation}
the metric $G$ is diagonal,
\begin{equation}\label{metric}
  G_{\mu\nu}=
  \begin{pmatrix}
  \theta_{(1)}^{-1}\I_2&0&0&\hdots \\
  0&\theta_{(2)}^{-1}\I_2&0&\hdots\\
  0&0&\theta_{(3)}^{-1}\I_2&\hdots\\
  \vdots&\vdots&\vdots&\ddots
  \end{pmatrix},
\end{equation}
where
\begin{equation}\label{mtrices}
  \ii\sigma_2=
  \begin{pmatrix}
  0&1\\
  -1&0
  \end{pmatrix}, \text{ and }
  \I_2=
  \begin{pmatrix}
  1&0\\
  0&1
  \end{pmatrix}.
\end{equation}

Then the solution is given by equation similar to \eqref{star_sol}
but with $\I(x)$ and $\sigma_{\alpha}(x)$ instead of respective
two-dimensional symbols.

% -----------------------------------------------------------------
\section{Discussions and Conclusions}

In this paper we considered the dynamics of interacting
noncommutative lumps.

The naive approach for the dynamics is obtained when one considers
the motion of the lumps as rigid structures and not their
deformations. The action in this approach is given by initial
classical action of the noncommutative model computed when all
``degrees of freedom'' except the positions of the lumps are
frozen. In this approximation one obtains that the lump pair
dynamics is described by a cup-shaped potential having minimum at
the origin, Gaussian decay at the infinity and an unstable
equilibrium at the distance $\sqrt{\theta\ln2}$.

The exact analysis of the interacting lump dynamics in the the
framework of the original noncommutative theory, however, refute
above approximation since it appears that in fact it is the shape
which is affected by the dynamics, while the motion of the centres
of the lumps is not.

It is interesting to note that the problem formulated in
noncommutative Yang--Mills model is reduced to one in finite
dimensional matrix model. Thus, in the case of two lumps starting
at the rest the exact description reduces to a $2\times2$ matrix
model. In particular we have that the U(1) part  of these model
has trivial dynamics, while the remaining SU(N) parts generally
exhibits stochastic behaviour.

The property of this dynamics that it does not affect the motion
along the line connecting the lumps appears to be anti-intuitive
to what one could expect from interaction of (quasi)particle
objects. Let us note that an analogous situation can be met in the
analysis of the vortex interaction in the applications to solid
state physics,
\cite{Papanicolaou:1991sg,Stratopoulos:1996,Papanicolaou1993},
where a behaviour similar to one of noncommutative lumps was
observed long ago.

It seems that these results can be easily generalised to the case
of lumps with arbitrary mutual Hilbert space polarisations not
related to the shifts along noncommutative space. The dynamics of
such lumps or branes does not differ qualitatively from the case
shifted ones, however, in this case we do not have simple physical
picture we can observe. However, from the point of view of
mathematical completeness this would be worthy to be considered,
and probably will be done in future research.

It seems that the interpretation in terms of branes when the
heights of the lumps have the meaning of  coordinates of the
0-brane in the direction transversal to the noncommutative brane
is the most natural. In this context it appears that the dynamics
of interacting branes affects only the motion in the transversal
directions in which D0-branes are stochastically ``bouncing''
around the noncommutative brane.

Translating the above said about lumps to the branes, we have
learnt that the dynamics of two interacting 0-branes is described
by U(2) M(atrix) model in the case when the branes do not move or
do not change their polarisations. If the branes are affected by
motion one needs a matrix model of higher dimension to describe
it.

So far we considered only the pair interaction of the lumps. It
also would be of interest to extend the analysis of the actual
paper to a greater number of the lumps, eventually to consider the
gas of lumps.

\subsection*{Acknowledgements}
I am grateful to Th. Tomaras, E. Kiritsis, K. Anagnostopoulos,
Nikos~Tetradis and Ciprian~Acatrinei for the friendly atmosphere
and hospitality during my stay in Crete where this work has been
done as well as to useful discussions and the interest paid to my
work. I would like to acknowledge the discussions on the vortex
interactions with Th. Tomaras. I benefited from useful discussions
at QFT dept of Steklov Mathematical Institute in Moscow.
Discussion with I.Ya. Aref'eva and P.B. Medvedev improved my
understanding the dynamics of the SU(2) matrix model.
% -----------------------------------------------------------------
\appendix
\section{Useful formula connecting the two dimensional representation
with other representations}

To improve the understanding of the paper we summarise here the formula
connecting the three main representations of the objects used in this
paper, Hilbert space operator, noncommutative functions (Weyl symbols)
and two-dimensional matrices.

The two-dimensional space $\hh_u$ for $u\neq0$ is the span of the
two vectors $\ket{u/2}=\e^{-\frac18 |u|^2}\e^{-\frac12
\bar{a}u}\ket{0}$ and $\ket{-u/2}=\e^{-\frac18 |u|^2}\e^{\frac12
\bar{a}u}\ket{0}$, where $\ket{0}$ is the oscillator vacuum state,
to which the lump operators $X_1$ and $X_2$ project to.

\begin{figure}
   \includegraphics[width=50mm]{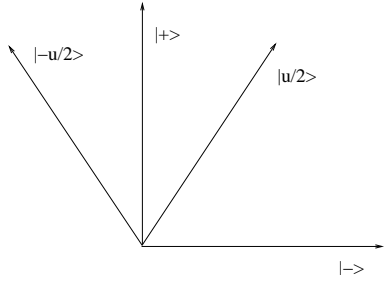}
   \caption{The orthogonal vectors $\ket{\pm}$ constructed from unimodular
   but nonorthogonal $\ket{\pm u/2}$.}
   \label{fig}
\end{figure}

The vectors $\ket{\pm u/2}$ have unity modules but are not
orthogonal. One can easily construct an orthonormal basis consisting
of vectors $\{\ket{+},\ket{-}\}$ given by eqs. (\ref{v+},\ref{v-})
(see picture~\ref{fig}).
% ----------------------------------------------------------------
% ----------------------------------------------------------------

An arbitrary Hermitian operator acting in this two-dimensional
subspace can be expanded in terms of ordinary Pauli matrices and unity
matrix,
\begin{equation}
   \sigma_0\equiv\I_2=
    \begin{pmatrix}
    1&0\\
    0&1
    \end{pmatrix}, \quad
   \sigma_1=
    \begin{pmatrix}
    0&1\\
    1&0
    \end{pmatrix},\quad
   \sigma_2=
     \begin{pmatrix}
    0&-\ii\\
    \ii&0
    \end{pmatrix},\quad
    \sigma_3=
     \begin{pmatrix}
    1&0\\
    0&-1
    \end{pmatrix},
\end{equation}
as follows,
\begin{equation}
  X=X^0+X^\alpha \sigma_\alpha,
\end{equation}
where $X^\alpha$ are computed as,
\begin{equation}
  X^\alpha=\frac12 \tr X\sigma_\alpha.
\end{equation}

% ----------------------------------------------------------------
\begin{figure}
   \includegraphics[width=50mm]{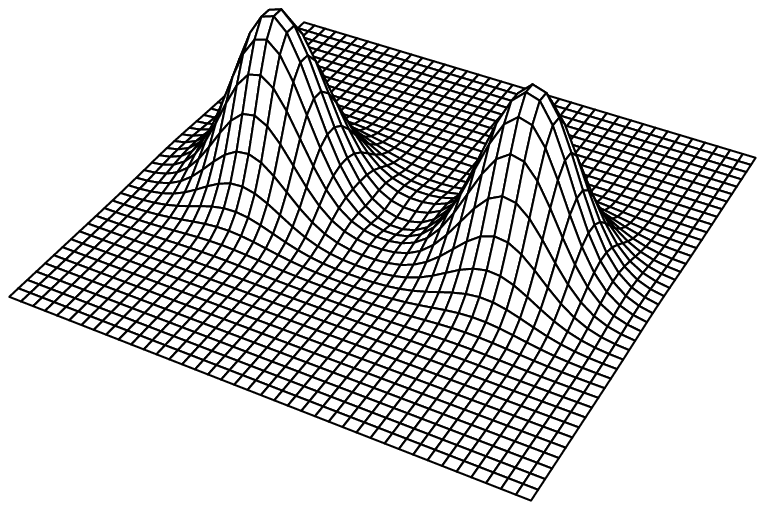}
   {\hspace{-1cm}$\sigma_0(\bar{z},z)$}
   \includegraphics[width=50mm]{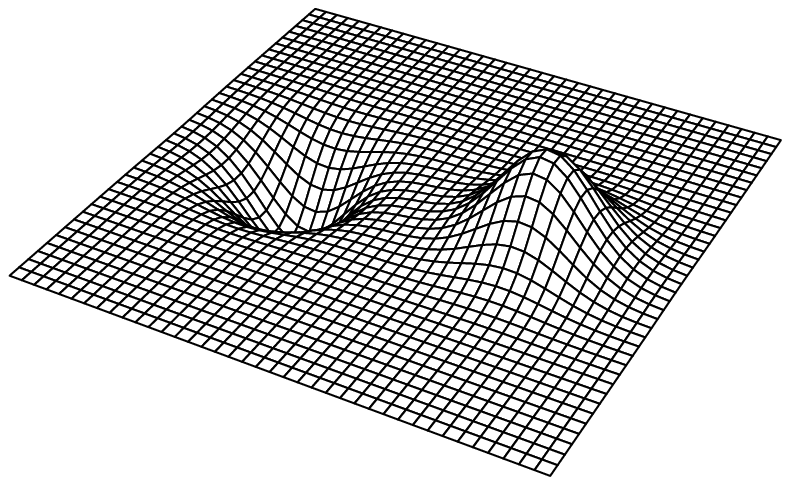}
   {\hspace{-1cm}$\sigma_1(\bar{z},z)$}
   \includegraphics[width=50mm]{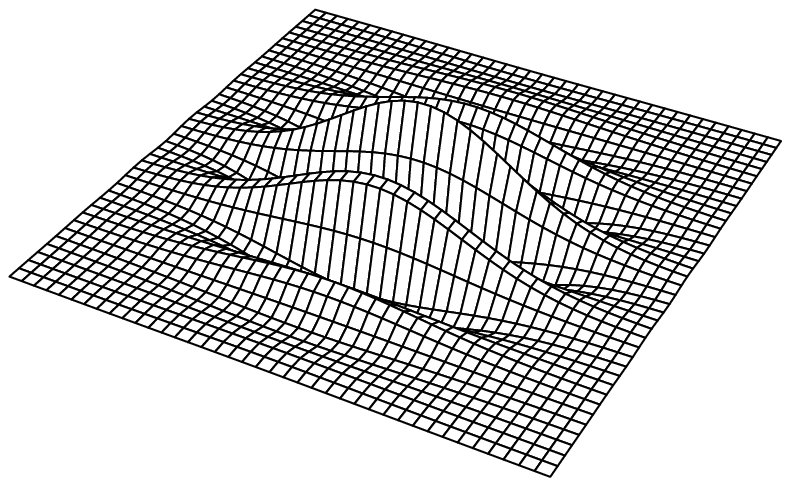}
   {\hspace{-1cm}$\sigma_2(\bar{z},z)$}
   \includegraphics[width=50mm]{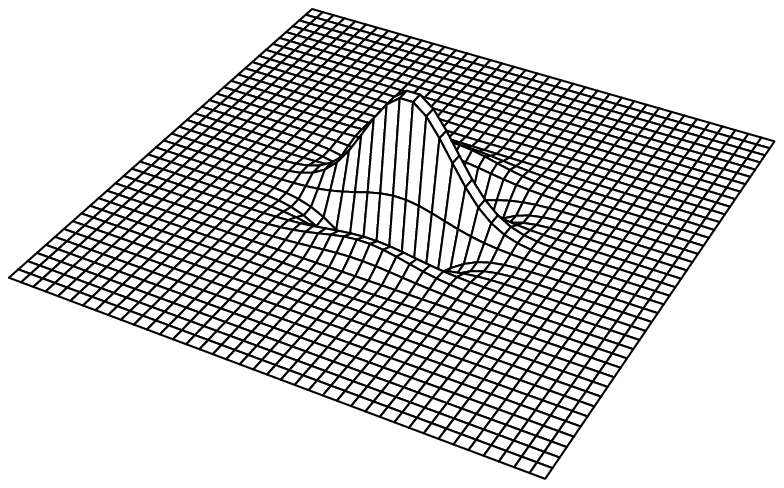}
   {\hspace{-1cm}$\sigma_3(\bar{z},z)$}
   \caption{Plots of the profiles of the functions
   $\sigma_{0,1,2,3}(\bar{z},z)$.}
   \label{sigmas_pl}
\end{figure}
% ----------------------------------------------------------------

From the other hand, as operators over the Hilbert space the
two-dimensional unity matrix\footnote{Which is the projector to
the two dimensional subspace $\hh_u$ of the Hilbert space.} and
Pauli matrices can be expressed as noncommutative functions
through their Weyl symbols.

The Weyl symbols of operators with bounded square-trace to which
undoubtedly belong $\I_2$ and $\sigma_\alpha$ can be found by a
direct formula,
\begin{equation}
  X\sim \int \dd \bar{k}\dd k\, \e^{\ii(\bar{k}z+k\bar{z})}\tr
  X\e^{-\ii(\bar{k}a+k\bar{a})}.
\end{equation}

Technically,  one can write the matrices in the (nonorthogonal)
basis of $\ket{\pm u/2}$ and use the Weyl symbols for the
following operators,
\begin{equation}\label{symbols}
\begin{split}
  \ket{u/2}\bra{u/2}& \sim 2\e^{-2|z-u/2|^2},\\
  \ket{-u/2}\bra{-u/2}& \sim 2\e^{-2|z+u/2|^2},\\
  \ket{u/2}\bra{-u/2}& \sim 2\e^{-2|z|^2+(\bar{z}u-z\bar{u})},\\
  \ket{-u/2}\bra{u/2}& \sim 2\e^{-2|z|^2-(\bar{z}u-z\bar{u})},
\end{split}
\end{equation}
which can be easily computed.

The sigma-matrices are expressed in the nonorthogonal basis of
$\ket{\pm u/2}$ as follows,
\begin{subequations}\label{nonort_sig}
\begin{align}
  \sigma_1&=\frac{1}{\sqrt{1-\e^{-|u|^2}}}
  (\ket{u/2}\bra{u/2}-\ket{-u/2}\bra{-u/2}), \\
  \sigma_2&=\frac{1}{\sqrt{1-\e^{-|u|^2}}}
  (\ket{u/2}\bra{-u/2}-\ket{-u/2}\bra{u/2}), \\
  \sigma_3&=-\frac{\e^{-\frac12 |u|^2}}{1-\e^{-|u|^2}}
  (\ket{u/2}\bra{u/2}+\ket{-u/2}\bra{-u/2})+\\ \nonumber
  &\frac{1}{1-\e^{-|u|^2}}
  (\ket{u/2}\bra{-u/2}+\ket{-u/2}\bra{u/2}),
\intertext{and, finally}
  \sigma_0&=\frac{1}{1-\e^{-|u|^2}}
  (\ket{u/2}\bra{u/2}+\ket{-u/2}\bra{-u/2})+\\ \nonumber
  &\frac{\e^{-\frac12 |u|^2}}{1-\e^{-|u|^2}}
  (\ket{u/2}\bra{-u/2}+\ket{-u/2}\bra{u/2})
\end{align}
\end{subequations}

Inserting \eqref{symbols} into \eqref{nonort_sig} one finds immediately
the functions \eqref{sigmas} of the third section.

The plots of functions $\sigma_\alpha(\bar{z},z)$ can be seen on the
fig.\ref{sigmas_pl}.

% ----------------------------------------------------------------
% \bibliographystyle{utphys}
% \bibliography{noncom}
% ----------------------------------------------------------------
\providecommand{\href}[2]{#2}\begingroup\raggedright\endgroup

\end{document}